\documentclass[prb,twocolumn,showpacs,preprintnumbers,amsmath,amssymb]{revtex4}
\usepackage{graphicx}% Include figure files
\usepackage{dcolumn}% Align table columns on decimal point
\usepackage{bm}% bold math

\begin{document}

%\preprint{APS/123-QED}

\title{Photoluminescence spectrum of an interacting two-dimensional
       electron gas at $\nu=1$}

\author{R.L. Doretto}
  \email{doretto@ifi.unicamp.br}
\author{A.O. Caldeira}%
% \email{caldeira@ifi.unicamp.br}
\affiliation{Departamento de F\'{\i}sica da Mat\'eria Condensada,
             Instituto de F\'{\i}sica Gleb Wataghin,
             Universidade Estadual de Campinas,
             Cep 13083-970, Campinas-SP, Brazil}

\date{\today}% It is always \today, today,

\begin{abstract}
We report on the theoretical photoluminescence spectrum of an
interacting two-dimensional electron gas at filling factor one
($\nu=1$). We considered a model similar to the one adopted to study
the X-ray spectra of metals and solved it analytically using the
bosonization method previously developed for the two-dimensional
electron gas at $\nu=1$. We calculated the emission spectra of the
right and the left circularly polarized radiations for the situations
where the distance between the two-dimensional electron gas and
the valence band hole is smaller and greater than the magnetic
length. For the former, we showed that the polarized
photoluminescence spectra can be understood as the recombination
of the so-called excitonic state with the valence band hole
whereas, for the latter, the observed emission spectra can be
related to the recombination of a state formed by a spin down
electron bound to $n$ spin waves. This state seems to be a good
description for the quantum Hall skyrmion.
\end{abstract}

\maketitle

\section{Introduction}

Among all the strongly correlated electron systems, the
two-dimensional electron gas (2DEG) at filling factor one ($\nu =
1$) has received a great deal of attention
lately.\cite{zyun,perspectives}  The
ground state of this system, the so-called quantum Hall
ferromagnet, is formed by a spin polarized state, where all
electrons completely fill the spin up lowest Landau level. The
neutral elementary excitations are described as spin waves or {\it
magnetic excitons} \cite{kallin} while the low-lying charged one
as a {\it charged spin texture} known as the quantum Hall skyrmion.
\cite{sondhi}

The analysis of the polarized photoluminescence spectrum of the
2DEG around $\nu=1$ allows us to study the charged excitations of
the system in the presence of a valence band hole. An interesting
aspect is that the emission spectrum of the right circularly
polarized (RCP) radiation is quite distinct from the left
circularly polarized (LCP) one.\cite{plentz} It has been observed
that while the RCP spectrum is formed by only one sharp peak, the
peak of the LCP spectrum is asymmetric with an additional spectral
weight on the low-energy side. The existence of a low-energy tail
in the LCP spectrum is related to the spin wave excitation which
is left in the system after the recombination process.
\cite{cooper} In both cases, the energy of the peaks continuously
decreases as the filling factor changes from $\nu<1$ to $\nu>1$.
Here, the estimated distance $d$ between the 2DEG and the valence
band hole was $d<l$, where $l = \sqrt{\hbar c/(eB)}$ is the
magnetic length. The distance $d$ is an important parameter for
the system as new features have been observed in the
photoluminescence spectrum when the distance $d$ increases.

Osborne {\it et al.} \cite{osborne} performed similar measurements
in a single quantum well sample applying an electric field
(perpendicular to the two-dimensional electron system), which
polarizes the electrons and the holes to opposite sides of the
quantum well, increasing the value of the distance $d$. For low
intensity electric field, the spectrum has the same
characteristics as the ones observed in Ref. \onlinecite{plentz}. As the
electric field increases, the energies of the peaks decrease and a
discontinuity appears at $\nu=1$, namely, there is a blue shift of
the energy of the peaks as the filling factor varies from
$\nu<1$ to $\nu>1$. Furthermore, the width of the LCP spectral
line on the low-energy side also increases. Those features are
enhanced when the photoluminescence spectrum of double quantum
well sample is analyzed. In this case, the distance between the
centers of the two quantum wells is $d>l$. It has been observed
that the spectra related to the electron-hole recombination in the
same (direct) and between distinct (indirect) quantum wells are
similar to the data recorded for the single quantum well sample in
the limit of low and high electric fields, respectively. In
particular, the increase of the width of the low-energy tail of
the LCP spectrum is now more evident and the RCP spectrum also
presents this behavior, differently from the $d < l$ case.

The discontinuity in the photoluminescence spectrum described
above for a high applied electric field is understood as a change
in the nature of the {\it photoexcited ground state}
as the filling factor changes from $\nu < 1$ to $\nu > 1$.\cite{cooper}
This is defined as the state of the system after the photoexcitation
and thermal relaxation processes which will recombine with the
valence band hole.\cite{osborne}

The nature of the photoexcited ground state is also strongly
related to the distance $d$ between the 2DEG and the valence band
hole. For instance, at $\nu=1$, Cooper and Chklovskii
\cite{cooper} showed that, if $d < l$, the photoexcited ground
state is formed by the completely filled spin up lowest Landau
level plus a spin down electron bound to the valence band hole,
the so-called {\it excitonic state}. In this case, the quantum
Hall ferromagnet is inert after the photoexcitation as the Coulomb
interaction between the spin down electron and the hole is
stronger than the electron-electron interaction within the 2DEG.
On the other hand, for larger values of the distance $d$, Osborne
{\it et al.} \cite{osborne} suggested that the photoexcited ground
state should be described by a quantum Hall skyrmion (the charged
excitation of the 2DEG at $\nu=1$) bound to the valence band hole,
i.e., a {\it skyrmion-hole} state. Only the recombination of such
state can leave a spin wave excitation in the system and therefore
we could understand the existence of low-energy tails in both RCP
and LCP spectra.

The aim of this paper is to calculate the form of the spectral
line of the polarized photoluminescence spectra of the interacting
two-dimensional electron gas at $\nu = 1$, in the limit of small
($d < l$) and large ($d > l$) separations between the 2DEG and the
valence band hole. {\it We will concentrate only on the filling factor
equal to one}. Our model is similar to the ones considered by
Schotte and Schotte to study the X-ray spectra of metals
\cite{schotte} and by Westfahl Jr. {\it et al.} to study the X-ray
edge problem for the 2DEG under a perpendicular magnetic field
when $\nu\gg 1$.\cite{harry2} We will show that the form of the
spectral line can be analytically calculated using the
bosonization method for the two-dimensional electron gas at $\nu =
1$ presented in Ref. \onlinecite{doretto}, and that the results are
qualitatively in agreement with the experimental data of Plentz
{\it et al.} \cite{plentz} and Osborne {\it et al.}
\cite{osborne}

The paper is organized as follows. In the next section, we will
present our model to describe the photoluminescence experiment.
In Sec. III, we will calculate
the form of the spectral line for both polarizations, for the case $d <
l$, considering that the photoexcited ground state is given by the
excitonic state, whereas in Sec. IV, we will do the same analysis for
the $d > l$ case, assuming that the skyrmion-hole state is the photoexcited
ground state. Finally, in Sec. V, we will present a summary of our
results.

\section{The model}

\begin{figure}[t]
\centerline{\includegraphics[height=4.5cm]{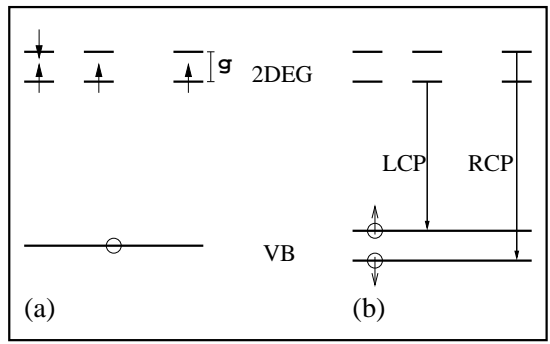}}
\caption{\label{esquema}{Schematic representation of the
    photoluminescence experiment in the 2DEG at $\nu=1$: (a) the
    valence band hole created after the photoexcitation and the extra
    electron added to the lowest Landau level of the 2DEG; (b) the
    emission of the right (RCP) and left (LCP) circularly polarized
    radiations. VB is the valence band and g is the Zeeman energy.}}
\end{figure}

The photoluminescence experiment for the 2DEG at $\nu=1$ is
schematically described in Fig. \ref{esquema}. After the
photoexcitation, a hole is created in the valence band and one
electron is created in the spin down lowest Landau level [Fig.
\ref{esquema} (a)]. There are two possible recombination processes
as illustrated in Fig. \ref{esquema} (b). If a spin down electron
recombines with a spin $-3/2$ valence band hole, the change of the
$z$ component of the total spin of the system is $\Delta S^Z = +1$
and therefore the emitted radiation is right circularly polarized.
On the other hand, if a spin up electron recombines with a spin
$3/2$  valence band hole, $\Delta S^Z = -1$, and hence the emitted
radiation is left circularly polarized.

We will consider that the distance between the planes of the 2DEG and
the valence band hole is $d$. Following \cite{schotte,harry2}, 
we will analyze the system before and after the recombination process.

The Hamiltonian of the system prior to the recombination is given by
\begin{equation}
\label{hi}
     \mathcal{H}_i = \mathcal{H}^e +\mathcal{H}_0^h
     + \mathcal{H}_{int}^{e-h}.
\end{equation}
Here, $\mathcal{H}^e$ is the Hamiltonian of the interacting
two-dimensional electron gas at $\nu \approx 1$ with all the
electrons restricted to the lowest Landau level. In the Landau
level basis, it can be written as [see Ref. \onlinecite{doretto} for
details]
\begin{eqnarray}
\nonumber
\mathcal{H}^e &=& -\frac{1}{2}g\sum_{\sigma}\sum_{m}\sigma
                   c^{\dagger}_{m\,\sigma}c_{m\,\sigma}  \\
\nonumber && \\ \label{he}
   && + \frac{1}{2}\sum_{\sigma,\sigma'}\int \frac{d^2k}{4\pi^2}\;
          V(\mathbf{k})\rho_{\sigma}(\mathbf{k})\rho_{\sigma'}(\mathbf{-k})
\end{eqnarray}
where $c^{\dagger}_{m\,\sigma}$ is a fermionic operator, which
creates a spin $\sigma$ electron in the lowest Landau level with
guiding center $m$. The density operator of spin $\sigma$
electrons is given by
\[
\rho_{\sigma}(\mathbf{k}) = e^{-|lq|^2/2}\sum_{m,m'}
      G_{m,m'}(lk)c^{\dagger}_{m\,\sigma}c_{m'\,\sigma},
\]
with the function $G_{m,m'}(x)$ defined in the appendix
\ref{funcaog}. $g=g^*\mu _BB > 0$ is the Zeeman energy, where
$g^*$ is the effective electron gyromagnetic factor in the host
semiconductor, $\mu_B$ is the Bohr magneton and $B$ the external
magnetic field. Finally, $ V(k) = 2\pi e^2/(\epsilon k)$ is the
Fourier transform of the Coulomb potential in two-dimensions
and $\epsilon$ is the dielectric constant of the semiconductor.

The valence band hole is described by the Hamiltonian,
\[
\mathcal{H}_0^h = \hbar w_0h^{\dagger}h,
\]
where the operator $h^{\dagger}$ creates a hole in the valence
band with energy $-\hbar w_0$ (measured from the chemical
potential). We assume that the hole wave function is localized at
the origin of the system of coordinates and that it has $s$-wave
symmetry.

The interaction between the 2DEG and the valence band hole is
given by a contact potential at the origin
\[
   \mathcal{H}_{int}^{e-h} =
   V_0\sum_{\sigma}\Psi^{\dagger}_{\sigma}(0)\Psi(0)_{\sigma}h^{\dagger}h,
\]
where $\Psi^{\dagger}_{\sigma}(\mathbf{r})$ is the fermionic field
operator at the origin
\begin{equation}
\label{fermionfield}
\Psi_{\sigma}^{\dagger}(\mathbf{r})
          = \sum_{m}\frac{1}{\sqrt{2\pi l^2}}e^{-|r|^2/4l^2}
             G_{0,m}(-ir^*/l)c^{\dagger}_{m\,\sigma}.
\end{equation}
For $\mathbf{r}=0$, the function $G_{0,m}(-ir^*/l)$ vanishes
except for $m=0$. This condition simplifies our
calculations. Since the angular momentum is conserved during the
recombination process (the electron-hole interaction potential is
spherically symmetric), only electrons with guiding center $m=0$
will recombine with the valence band hole.

After the recombination process, we end up with only
the two dimensional electron system at $\nu = 1$. Therefore the Hamiltonian 
of the
system is given by Eq. (\ref{he}),
\begin{equation}
\mathcal{H}_f = \mathcal{H}^e.
\end{equation}

In order to describe the emission spectrum at $T=0$, we need to calculate
the transition rate ($\hbar=1$),
\begin{equation}
\label{taxa}
W(w) \propto \sum_{f} |\langle f|\Psi_{\sigma}(0)|i\rangle|^2
                 \delta(w-(w_0 + E_i - E_f)),
\end{equation}
where $|f\rangle$ is the set of all final states of the system
after the recombination process with energy $E_f$ and $|i\rangle$
is the photoexcited ground state with energy $E_i$. Substituting
Eq. (\ref{fermionfield}) in the expression (\ref{taxa}), the
transition rate can be written as
\begin{eqnarray}
\nonumber W(w) %&\propto& \sum_{m}\sum_{f} |\langle
%f|c_{m\;\sigma}|i\rangle|^2
%                 \delta(w-(E_f - E_i - w_0)) \\
%\nonumber &&\\\nonumber
         &\propto& \sum_{f} |\langle f|c_{m=0\;\sigma}|i\rangle|^2
                 \delta(w-(w_0 + E_i - E_f)) \\
\nonumber &&\\\nonumber
         &\propto& \sum_{f}\int^{\infty}_{-\infty} dt\;
         e^{-i(w - w_0 - E_i + E_f)t}\;
         \langle i|c^{\dagger}_{0\;\sigma}|f\rangle \\
\nonumber && \\ \nonumber
         && \times\langle f|c_{0\;\sigma}|i\rangle \\
\nonumber &&\\ \nonumber
         &\propto& Re\int^{\infty}_0 dt\; e^{-i(w -w_0)t}\;
         \langle i|e^{i\mathcal{H}_it} c^{\dagger}_{0\;\sigma}
         e^{-i\mathcal{H}_ft}c_{0\;\sigma}|i\rangle
\\ \label{taxaauxiliar} &&\\ \nonumber
         &\propto& Re\int^{\infty}_0 dt\; e^{-i(w -w_0 - E_i)t}\;
         \langle i|c^{\dagger}_{0\;\sigma}
         e^{-i\mathcal{H}_ft}c_{0\;\sigma}|i\rangle.
\\ \label{taxa1}
\end{eqnarray}
Here, $\sigma = \uparrow$ and $\downarrow$ give the LCP and RCP
transition rates, respectively.

In Refs. \onlinecite{schotte} and \onlinecite{harry2}, the         
transition rate is calculated
from the expression (\ref{taxaauxiliar}) using the respective
bosonization methods. However, since we do not have an expression
for the fermionic field operator in terms of the annihilation and
creation bosonic operators and it is not possible to write down
the bosonic form of the Hamiltonian $\mathcal{H}_i$, which
describes the 2DEG at $\nu \approx 1$, Eq. (\ref{taxaauxiliar})
can not be properly written in the bosonic form. Therefore, as our
bosonization method for the 2DEG at $\nu=1$ presents some
limitations, we will employ Eq. (\ref{taxa1}) in our calculations.
All those points will become clearer in the next section.

Since we will not consider the Hamiltonian $\mathcal{H}_i$
explicitly, the energy $E_i$ will be treated as a parameter of our
model. Following this scheme, we are not able to determine the
energy of the photoexcited ground state but once we have the form
of the initial state $|i\rangle$ we can {\it analytically} calculate
the emission spectrum.

\section{The excitonic initial state}

First of all, we will analyze the case where the distance between
the 2DEG and the valence band hole is small, namely, $d<l$. We will
compare our results with the ones of Cooper and Chklovskii
\cite{cooper} and with the experimental data of Plentz {\it el al.}
\cite{plentz} in
order to check that our scheme is quite reasonable to describe
qualitatively the photoluminescence spectrum in this limit.
We will consider {\it only} the case $\nu=1$. 

As has been pointed out in \cite{cooper}, in this limit the
photoexcited ground state can be described by the {\it excitonic
state}, which is formed by the quantum Hall ferromagnet,
$|FM\rangle$, plus one electron of spin down in the lowest Landau
level and one valence band hole. This state can be written as
\begin{equation}
\label{excitonic}
|i\rangle = c^{\dagger}_{m\;\downarrow}|FM\rangle.
\end{equation}
Since the electron-hole interacting potential is spherically symmetric and
the valence band hole is assumed to be localized at the origin, we
can choose $m=0$ in the above expression.

Substituting Eq. (\ref{excitonic}) in the LCP transition rate
expression [Eq. (\ref{taxa1})], we have
\begin{eqnarray}
\nonumber W^{0}_{LCP}(w) &\propto& Re\int^{\infty}_0 dt\;
        e^{-i(w - w_0 - E_i)t}\\
\nonumber && \\ \label{taxa2}
        &&\times\underbrace{\langle 
FM|c_{0\;\downarrow}c^{\dagger}_{0\;\uparrow}
        e^{-i\mathcal{H}_ft}c_{0\;\uparrow}c^{\dagger}_{0\;\downarrow}
        |FM\rangle}_{\mathcal{F}_{LCP}(t)}.
\end{eqnarray}

The function $\mathcal{F}_{LCP}(t)$ can be easily calculated using
the bosonization method for the two-dimensional electron gas at $\nu=1$ 
introduced in Ref. \onlinecite{doretto}. It was shown that the Hamiltonian of the
interacting two-dimensional electron gas at $\nu=1$,
$\mathcal{H}^e$, can be mapped into an interacting two-dimensional
bosonic model
\begin{eqnarray}
\nonumber
\mathcal{H} &=& -\frac{1}{2}gN_{\phi} +
      \sum_{\mathbf{q}}w_{\mathbf{q}}b^{\dagger}_{\mathbf{q}}b_{\mathbf{q}}
\\ \nonumber && \\ \nonumber
    && +\frac{2}{\mathcal{A}}\sum_{\mathbf{k,p,q}}V(k)e^{-|lk|^2/2}
       \sin(\mathbf{k}\wedge\mathbf{p}/2)
\\ \nonumber && \\ \label{hintboso}
    && \times\sin(\mathbf{k}\wedge\mathbf{q}/2)
       b^{\dagger}_{\mathbf{k}+\mathbf{q}} 
b^{\dagger}_{\mathbf{p}-\mathbf{k}}
                         b_{\mathbf{q}}b_{\mathbf{p}},
\end{eqnarray}
where the boson operators $b$ are writing in terms of the fermionic
operators as
\begin{eqnarray}
\label{b}
   b_{\mathbf{q}} &=& \frac{1}{\sqrt{N_{\phi}}}
   e^{-|lq|^2/4}\sum_{m,m'}
   G_{m,m'}(-lq)c^{\dagger}_{m\,\uparrow}c_{m'\,\downarrow} \\ \nonumber
&&\\
\label{b+}
   b^{\dagger}_{\mathbf{q}} &=& \frac{1}{\sqrt{N_{\phi}}}
   e^{-|lq|^2/4}\sum_{m,m'}
   G_{m,m'}(lq)c^{\dagger}_{m\,\downarrow}c_{m'\,\uparrow},
\end{eqnarray}
which obey the canonical commutation relations
\begin{eqnarray}
[b^{\dagger}_{\mathbf{q}},b^{\dagger}_{\mathbf{q'}}] &=&
[b_{\mathbf{q}},b_{\mathbf{q'}}] = 0, \\ \nonumber
[b_{\mathbf{q}},b^{\dagger}_{\mathbf{q'}}] &=&
\delta_{\mathbf{q},\mathbf{q'}}.
\end{eqnarray}
The dispersion relation of the bosons $b$ is given by
\begin{equation}
\label{rpa}
w_{\mathbf{q}} = g + \frac{e^2}{\epsilon l}\sqrt{\frac{\pi}{2}}
           \left(1 - e^{-|lq|^2/4}I_0(|lq|^2/4)\right),
\end{equation}
where $I_0(x)$ is the modified Bessel function of the first
kind.\cite{grads} 
The state $b^{\dagger}_{\mathbf{q}}|FM\rangle$ corresponds to a
spin wave excitation with momentum $\mathbf{q}$ of the quantum
Hall ferromagnet and to a very well separeted quasiparticle-quasihole
pair, respectively for $|lq| < 1$ and $|lq| \gg 1$. This result is
exactly the one previously obtained by Kallin and
Halperin.\cite{kallin} Here, we will approximate $\mathcal{H}^e$ by the
noninteracting bosonic model, i.e.,
\begin{equation}
\label{ho}
      \mathcal{H}_f = \mathcal{H}^e = -\frac{1}{2}gN_{\phi} +
      \sum_{\mathbf{q}}w_{\mathbf{q}}b^{\dagger}_{\mathbf{q}}b_{\mathbf{q}}. 
\end{equation}

As pointed out in the previous section, our bosonization method
can be applied only to the 2DEG at $\nu=1$ and therefore it can
not be used to bosonize the Hamiltonian $\mathcal{H}_i$, which
describes the 2DEG at $\nu = 1$ with an extra electron ($\nu
\approx 1$). Therefore, we decide to keep the energy $E_i$ as a
parameter of our model. Despite the fact that we do not have an
expression for the fermionic field operators in terms of the
bosonic operators $b$, we can bosonize products of creation and
annihilation fermionic operators, such as
$c^{\dagger}_{0\;\downarrow}c_{0\;\uparrow}$. The obtained
expressions are presented in the appendix \ref{cc}. Based on the
above mentioned points, it is now clear why we need to choose Eq.
(\ref{taxa1}) instead of Eq. (\ref{taxaauxiliar}) to calculate the
transition rate and also why we need to write down the form of the
photoexcited ground state.

Substituting Eqs. (\ref{ho}), (\ref{cc1}) and (\ref{cc2}) in the expression
(\ref{taxa2}) and using the fact that $|FM\rangle$ is the vacuum for the
bosons $b$, the function $\mathcal{F}_{LCP}(t)$ reads
\begin{eqnarray}
\nonumber
    \mathcal{F}_{LCP}(t) &=& \sum_{\mathbf{q},\mathbf{q'}}
     \frac{e^{-|lq|^2/4-|lq'|^2/4}}{N_{\phi}}
     \langle 
FM|b_{\mathbf{q}}e^{-i\mathcal{H}_0t}b^{\dagger}_{\mathbf{q'}}|FM\rangle
\\\nonumber && \\\nonumber
     &=& 
\sum_{\mathbf{q},\mathbf{q'}}\frac{e^{-|lq|^2/4-|lq'|^2/4}}{N_{\phi}}
      \delta_{\mathbf{q},\mathbf{q'}}
     e^{-it(w_{\mathbf{q}} + E_f + E_{FM})},
\end{eqnarray}
where $E_{FM} = -gN_{\Phi}/2$ is the energy of the state
$|FM\rangle$. In the second step above, we use the
relation
\begin{equation}
\label{auxiliar}
e^{-it\sum_{\mathbf{q}}w_{\mathbf{q}}b^{\dagger}_{\mathbf{q}}b_{\mathbf{q}}}
    b^{\dagger}_{\mathbf{k}} =
    b^{\dagger}_{\mathbf{k}}\exp(-itw_{\mathbf{k}})
e^{-it\sum_{\mathbf{q}}w_{\mathbf{q}}b^{\dagger}_{\mathbf{q}}b_{\mathbf{q}}},
\end{equation}
which can be proved with the aid of the Baker-Hausdorff formula. 
\cite{mahan} Note that the terms $e^{-|lq|^2/4}$ in the
$\mathcal{F}_{LCP}(t)$ expression imply that only the long wavelength
excitations are important and therefore we can write the energy of the
bosons as
\begin{equation}
\label{expansaorpa}
w_{\mathbf{q}} \approx g + \epsilon_B(lq)^2/4,
\end{equation}
where $\epsilon_B = \sqrt{\pi/2}e^2/(\epsilon l)$ is a constant
related to the Coulomb energy $e^2/(\epsilon l)$ and $\epsilon$
is the dielectric constant of the host semiconductor. Changing the
sum over momenta into an integral,
\[
\frac{1}{\mathcal{A}}\sum_{\mathbf{q}} \rightarrow \int\frac{d^2q}{4\pi^2},
\]
where the area of the system is related to the magnetic length and to
the degeneracy of the lowest Landau level $N_{\phi}$ by $\mathcal{A} =
2\pi l^2N_{\phi}$, we can calculate the function
$\mathcal{F}_{LCP}(t)$ analytically, i.e.,
\begin{eqnarray}
\nonumber
\mathcal{F}_{LCP}(t) &=& 2\left(4 + t^2\epsilon_B^2\right)^{-1/2}
\exp\left(-it(g + E_{FM})\right) \\
\nonumber && \\ \label{flcp0}
&&\times\exp\left(-it\tan^{-1}(t\epsilon_B/2)\right).
\end{eqnarray}
Substituting the above expression in equation (\ref{taxa2}), we can
also solve the time integral analytically and obtain the
emission rate for the left circularly polarized radiation,
\begin{eqnarray}
\nonumber
  W^0_{lcp}(w) &\propto& \frac{\pi}{\epsilon_B}
                  sign\left(w^0_{lcp} - w\right)
\\ \nonumber \\ \label{lcp0}
  &&\times\exp\left(2(w-w^0_{lcp})/\epsilon_B\right),
\end{eqnarray}
where $w^0_{lcp} = w_0 + E_i - E_{FM} - g$ and $sign(x)$ is the
signal function.

\begin{figure}[t]
\centerline{\includegraphics[height=5.0cm]{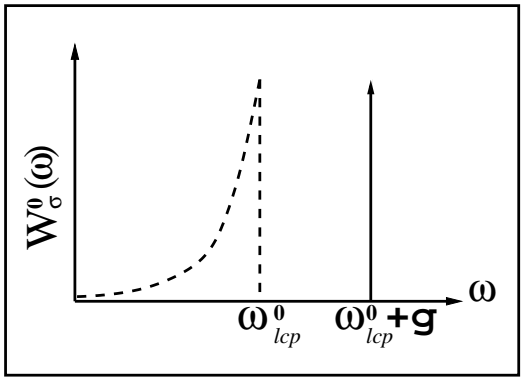}}
\caption{\label{espectros0}{Schematic representation of the
    emission spectra of the left (dashed line) and right (solid line)
    circularly polarized radiation when the
    photoexcited ground state is given by Eq. (\ref{excitonic}).}}
\end{figure}

As we can see in Fig. \ref{espectros0} (dashed line), the LCP
spectrum is formed by a peak at $w = w^0_{lcp}$ with a low energy
tail, which agrees with the results previously derived in Ref.
\onlinecite{cooper} and with the experimental data of Ref.
\onlinecite{plentz}. The observed behavior is related to the spin wave
excitation which remains after the recombination process (see Fig.
\ref{esquema}). Here, we should mention that the linewidth of
the experimental spectrum is related to the disorder effects as
showed in Ref. \onlinecite{cooper}. Therefore, the low energy tail
accounts for the asymmetry observed in the experimental spectrum.

Following an analogous procedure, we can calculate the transition
rate for the RCP radiation. In this case, we have
\begin{eqnarray}
\nonumber
W^{0}_{RCP} &\propto& Re\int^{\infty}_0 dt\; e^{-i(w - w_0 - E_i)t}
\\ \nonumber \\
       &&\times\underbrace{\langle 
FM|c_{0\;\downarrow}c^{\dagger}_{0\;\downarrow}
        e^{-i\mathcal{H}_ft}c_{0\;\downarrow}c^{\dagger}_{0\;\downarrow}
        |FM\rangle}_{\mathcal{F}_{RCP}(t)}.
\end{eqnarray}
Using the expression of the operator
$c^{\dagger}_{0\;\downarrow}c_{0\;\downarrow}$ in terms of the
bosons [Eq. (\ref{cc4})], it is possible to show that
\begin{equation}
\mathcal{F}_{RCP}(t) = e^{-iE_{FM}t},
\end{equation}
which implies that the emission rate of the RCP radiation is
simply given by
\begin{eqnarray}
\label{rcp0}
  W^0_{RCP}(w) &\propto& \delta\left(w-(w^0_{lcp}+g)\right).
\end{eqnarray}
The above expression is schematically illustrated in 
Fig. \ref{espectros0} (solid line).
Based on this result, we can conclude that 
the RCP spectrum is formed only by a sharp peak, whose
energy is greater than the energy of the LCP peak. Differently
from the LCP spectrum, the form of the RCP spectral line is
not asymmetric. Again, our results are in qualitative agreement with
the experimental data.\cite{plentz,osborne}

Despite the fact that our model is not able to determine the
energy of the photoexcited ground state and therefore the energies
of the peaks of the LCP and RCP spectra, it captures the main
features of the experimental polarized photoluminescence spectra
reported in \cite{plentz,osborne}, where the estimated value of
the distance $d$ was smaller than the magnetic length. 

In the next section, we will extend our analysis to the $d > l$ case
in order to check the suggestion of  Osborne {\it et al.}
\cite{osborne}, which pointed the existence of skyrmions in the 2DEG 
prior to the recombination process. As far as we know, this is the
first time that the form of the spectral line of the polarized
photoluminescence spectrum of the 2DEG at $\nu=1$ is calculated,
including the presence of skyrmions in the system before the
recombination.  

\section{The skyrmion-hole initial state}

Now, we will consider that the distance between the 2DEG and the
valence band hole is $d > l$. In this case, the Coulomb
interaction between the spin down electron and the spin up
electrons is greater that the interaction between the former and
the valence band hole (remember that, at $\nu=1$, the radius of
the cyclotron orbit of each electron is equal to $l$). As a
result, the photoexcited ground state changes from the excitonic
state [Eq. (\ref{excitonic})] to a quantum Hall skyrmion bound to
the valence band hole (skyrmion-hole state). This scenario is also
corroborated by the numerical calculations of Portengen {\it et al.}
\cite{portengen}, who showed that the skyrmion-hole state is more
stable than the excitonic state when $d > l$. The presence of
skyrmions in the system prior to the recombination is also discussed
in \cite{wojs} and the references therein. 

The above modification in the nature of the photoexcited ground
state, in relation to the one considered in the last section, can
be easily accommodated in our model. It can be done considering
that the skyrmion (a charged excitation of the quantum Hall
ferromagnet with spin $S^z > 1/2$) can be described by a state formed
by a spin down
electron bound to a determined number of spin wave excitations. In
fact, this idea was suggested by Palacios and Fertig
\cite{palacios}, but only Oaknin {\it et al.} studied this model
in details.\cite{oaknin} It is also considered in our previous work.
\cite{doretto} 

If we assume that the skyrmion is formed by one spin down electron
bound to $n$ spin waves, we
can write down the photoexcited ground state as
\begin{equation}
   |i\rangle \sim \prod_{i=1}^nb^{\dagger}_{\mathbf{q}_i}
   c^{\dagger}_{m\,\downarrow}|FM\rangle,
\label{skyrmion}
\end{equation}
where $|lq_i| < 1$.
In the above expression, we considered the fact that, within our
bosonization method, the spin wave excitations of the interacting
two-dimensional electron gas at $\nu=1$ can be described
approximately as bosons.

Substituting Eq. (\ref{skyrmion}) in (\ref{taxa1}), the transition
rate for the left circularly polarized radiation reads
\begin{widetext}
\begin{eqnarray}
\nonumber
W_{LCP}(w) &\propto& Re\int^{\infty}_0 dt\; e^{-i(w - w_0 - E_i)t}
%\\ \nonumber && \\ \nonumber
%    && \times
        \underbrace{\langle FM|c_{m\;\downarrow}\prod_{i=1}^nb_{\mathbf{q}_i}
        c^{\dagger}_{0\;\uparrow}
        e^{-i\mathcal{H}_ft}c_{0\;\uparrow}
        \prod_{i=1}^nb^{\dagger}_{\mathbf{q}_i}c^{\dagger}_{m\;\downarrow}
        |FM\rangle}_{\mathcal{F}_{LCP}(t)}.
\\  \label{taxalcp}
\end{eqnarray}
Here, we will follow the same procedure of the previous section,
namely, the energy of the initial state $E_i$ will be treated as a
parameter and the Hamiltonian $\mathcal{H}_f$ will be given by Eq.
(\ref{ho}). As
$[b^{\dagger}_{\mathbf{q}},c^{\dagger}_{m\;\uparrow}]=0$, it is
possible to reorder the fermionic operators in equation
(\ref{taxalcp}) and, using the bosonic representation of the
product of fermionic operators
$c^{\dagger}_{m\;\sigma}c_{m\;\sigma'}$ [see Eqs. (\ref{cc1}) and
(\ref{cc2})], we have
\begin{eqnarray}
\mathcal{F}_{LCP}(t) &=&
\frac{1}{N_{\phi}}\sum_{\mathbf{k}}e^{-|lk|^2/2}G_{m,0}(lk)
               G_{0,m}(-lk)\exp\left(-it(\epsilon + w_{\mathbf{k}} + E_{FM})\right) 
\nonumber \\ && \nonumber \\
           && + \frac{1}{N_{\phi}}\sum_{i=1}^n e^{-|lq_i|^2/2}G_{m,0}(lq_i)
                G_{0,m}(-lq_i)\exp\left(-it(\epsilon + w_{\mathbf{q}_i} + E_{FM})\right),
\end{eqnarray}
where $\epsilon = \sum_{i=1}^nw_{\mathbf{q}_i}$ is the energy of the
spin waves present in (\ref{skyrmion}).
Again, we will change the sum over momenta into an integral and expand
$w_{\mathbf{k}}$ as in equation (\ref{expansaorpa}). Assuming that
$m=0$, the function
$G_{0,0}(x)=1$. As a result, the integral over momenta of the above
expression can be calculated analytically,
\begin{eqnarray}
\nonumber
\mathcal{F}_{LCP}(t) &=& 2(4 + t^2\epsilon_B^2)^{-1/2}
  \exp\left(-it(\epsilon + g + E_{FM})\right)
  \exp\left(-i\tan^{-1}(t\epsilon_B/2)\right)
\\ \nonumber && \\ \label{flcp}
&& + \frac{1}{N_{\phi}}\exp\left(-it(\epsilon + E_{FM})\right)
     \sum_{i=1}^n\exp\left( - itw_{\mathbf{q}_i} - |lq_i|^2/2 \right).
\end{eqnarray}
Now, substituting Eq. (\ref{flcp}) in (\ref{taxalcp}), we can
analytically calculate the temporal integral and show
that the emission rate for the LCP radiation is simply given by
\begin{eqnarray}
W_{LCP}(w) &\propto& \frac{2\pi}{\epsilon_B}
                  sign\left(w_{lcp}-w\right)
                  \exp\left(2(w-w_{lcp})/\epsilon_B\right) %\nonumber \\
% && \nonumber \\ &&  
           + \frac{1}{N_{\phi}}\sum_{i=1}^n\exp\left(-|lq_i|^2/2\right)
           \delta\left(w-(w_{lcp} + g - w_{\mathbf{q}_i})\right),
\label{lcp}
\end{eqnarray}
\end{widetext}
where $w_{lcp} = w_0 + E_i - E_{FM} - g - \epsilon$.

\begin{figure}[t]
\centerline{\includegraphics[height=5.0cm]{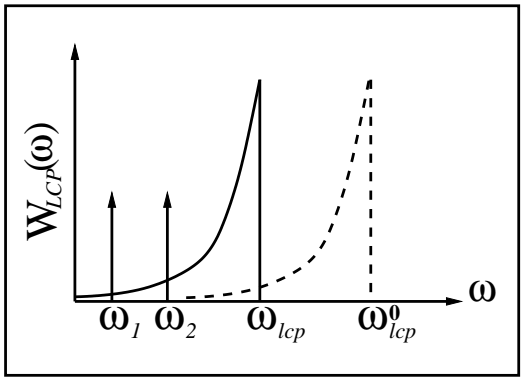}}
\caption{\label{espectros}{Schematic representation of the
    emission spectra of the left circularly polarized radiation when the
    photoexcited ground state is given by Eq. (\ref{skyrmion}) with $n=2$
    (solid line) and by Eq. (\ref{excitonic}) (dashed line).}}
\end{figure}

As illustrated in Fig. \ref{espectros} for the case $n=2$ (solid
line), $W_{LCP}$ is formed by a main peak localized at $w_{lcp}$ with a low-energy
tail, features which were observed in $W^0_{LCP}$. It is possible
to estimate the energy $E_i$ of the photoexcited ground state
$|i\rangle$ and show that $w_{lcp} < w^0_{lcp}$. In fact, writing
the energy of the state (\ref{excitonic}) as $E^0_i$ and the spin
waves-spin down electron binding energy as $E_b$, we have $E_i
\approx E^0_i + \epsilon - E_b$ and therefore $w^0_{lcp} -
w_{lcp} = E_b$. From this analysis, we can conclude that there is
a redshift in the energy of the main peak of $W_{LCP}$ in relation
to the one of $W^0_{LCP}$ and that this redshift is given by the
binding energy of the spin waves-spin down electron.

Moreover, the LCP spectrum also has a set of secondary sharp peaks at $w_i =
w_{lcp}+g-w_{\mathbf{q}_i}$ whose intensities are smaller than the
intensity of the main peak. Notice that the redshift of the energy of
each secondary peak $w_i$, in relation to the main one of $W_{LCP}$, 
is mainly given by the energy of the spin wave with momentum $\mathbf{q}_i$,
which is present in the state (\ref{skyrmion}).
The presence of those small peaks will increase the spectral weight
of $W_{LCP}(w)$ on the low energy side.

\begin{figure}[t]
\centerline{\includegraphics[height=5.0cm]{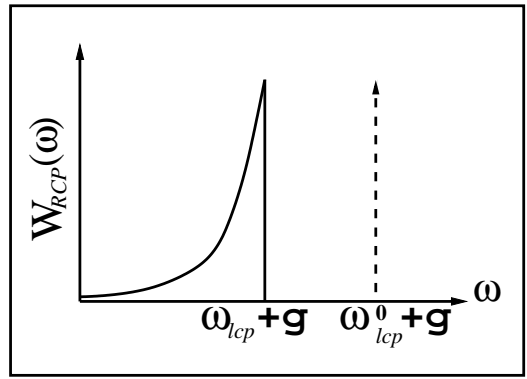}}
\caption{\label{espectrorcp}{Schematic representation of the
    emission spectra of the right circularly polarized radiation when the
    photoexcited ground state is given by Eq. (\ref{skyrmion})
    (solid line) and by Eq. (\ref{excitonic}) (dashed line).}}
\end{figure}

Finally, using the same approach, we can calculate the emission
rate for the right circularly polarized radiation. In this case,
Eq. (\ref{taxa1}) reads
\begin{widetext}
\begin{eqnarray}
\nonumber
  W_{RCP}(w) &\propto& Re\int^{\infty}_0 dt\; e^{-i(w - w_0 - E_i)t}
%\\ \nonumber && \\ \nonumber
%       && \times
        \underbrace{\langle FM|c_{m\;\downarrow}\prod_{i=1}^n b_{\mathbf{q_i}}
        c^{\dagger}_{0\;\downarrow}
        e^{-i\mathcal{H}_ft}c_{0\;\downarrow}
        \prod_{i=1}^n b^{\dagger}_{\mathbf{q_i}}c^{\dagger}_{m\;\downarrow}
        |FM\rangle}_{\mathcal{F}_{RCP}(t)}.
\\\label{taxarcp}
\end{eqnarray}
After a length calculation [see Appendix \ref{apendice:rcp}], it is
possible to show that $\mathcal{F}_{RCP}(t)$ can be written as  
\begin{eqnarray}
\nonumber
\mathcal{F}_{RCP}(t) &\approx& \delta_{m,0}\left(1 - \frac{2n}{N_\phi}
 + \frac{n(n-1)}{N^2_\phi}\right) 
        \exp\left(-it(E_{FM} + \epsilon)\right)
        -2i\frac{n}{N_\phi}\exp\left(-it(E_{FM} + \epsilon)\right)\frac{1}{t\epsilon_B - 2i}
\\ \nonumber && \\ \label{frcp}          
        && + \frac{1}{N_\phi^2}\sum_{j,l\not=j}\exp\left(-|l(\mathbf{q_j} -
             \mathbf{q_l})|^2/2\right)
             \exp\left(-it(\epsilon + E_{FM} - w_\mathbf{q_j} + w_\mathbf{q_l})\right) 
\end{eqnarray}
Substituting $\mathcal{F}_{RCP}(t)$ in (\ref{taxarcp}), neglecting the
terms of order $1/N^2_\phi$ and after the temporal integration, we have
\begin{eqnarray}
\label{rcp}
W_{RCP}(w) &\propto& \delta\left(w- w_{lcp} - g \right)
+ \frac{4\pi n}{\epsilon_BN_{\phi}}sign\left(w_{lcp} + g - w\right)
              \exp\left(2(w - w_{lcp} - g)/\epsilon_B\right) + \mathcal{O}(1/N^2_\phi).
\end{eqnarray}
\end{widetext}
Equation (\ref{rcp}) is schematically illustrated in Fig.
\ref{espectrorcp} (solid line). Differently from $W^0_{RCP}(w)$,
$W_{RCP}(w)$ has a sharp peak at $w_{lcp}+g$ in addition to a
low-energy tail [second term of Eq.(\ref{rcp})], which implies
that the RCP spectral line is also asymmetric. Here, the recombination
of a spin down electron with the valence band hole also leaves spin
waves in the 2DEG as the photoexcited ground state is formed by 
a spin down electron bound to $n$-spin waves. We can also observe a
redshift in the energy of this peak in relation to the one of
$W^0_{RCP}(w)$, which is related to the spin waves-spin down
electron binding energy.

The expressions (\ref{lcp}) and (\ref{rcp}) are in good qualitative
agreement with the experimental photoluminescence spectra
reported by Osborne {\it et al.} \cite{osborne} for a single
quantum well sample under high applied electric field and for the
indirect process in double quantum well samples. As pointed out in
Sec. I, a redshift of the energies of the main
peaks of the LCP and RCP spectra was observed in relation to the results
obtained for the single quantum well samples under a low electric
field. Our analysis showed that this redshift is related to the
change in the nature of the photoexcited ground state from the
excitonic state to the skyrmion-hole one. In particular, this
redshift is proportional to the binding
energy of the spin waves to the spin down electron, which form 
the skyrmion-hole state. Moreover, as the distance $d$ increases, 
the width of the low energy tail of the LCP spectrum also increases
and now the RCP spectrum is also asymmetric. The expression
(\ref{lcp}) has a low energy tail in addition to a set of secondary peaks
which account for the increase of the width of the low energy
tail of the LCP spectral line. Note that the spectral line
(\ref{rcp}) is also asymmetric.

Therefore, the polarized photoluminescence spectrum of the
two-dimensional electron gas at $\nu=1$, in the limit where $d > l$,
can be described by the recombination of a
state formed by a spin down electron 
bound to $n$ spin wave excitations with the valence band hole as 
suggested by Osborne {\it et al.}\cite{osborne}

Furthermore, we can also conclude that this state is a
good candidate to describe the elementary charged excitation of
the quantum Hall ferromagnet. Remember that, in the limit of larger
$d$, the effect of the hole over the two-dimensional electron gas
is smaller than in the case $d < l$. As a consequence, after the addition
of the spin down electron, the system probably relax to this charged
excited state before the recombination.

\section{Summary}

We studied the polarized photoluminescence spectrum of the
two-dimensional electron gas at $\nu=1$ in the limit of small and
large separation between the 2DEG and the valence band hole.

In order to do that, we considered a model analogous to the one
adopted to study the X-ray spectra of metals. The RCP and LCP
spectral lines were calculated using a previously developed
bosonization method for the 2DEG at $\nu=1$. 

For small distances $d$, we use the fact that the photoexcited
ground state can be described by an excitonic state and showed
that the LCP spectrum is formed by a peak with a low-energy tail,
which is related to the spin wave excitation left over after the
recombination process takes place, and that the RCP spectral line
is formed by a sharp peak.

For large distances $d$, we assume that the photoexcited ground
state is given by the skyrmion-hole state, which was described by
a spin down electron bound to $n$ spin waves. We showed that, in
addition to the main peak, the LCP spectrum also presents a set of
secondary peaks which accounts for the increase of the width of
the experimental spectral line on the low-energy side.
Furthermore, there is a redshift in the energy of the main peak,
which is closely related to the binding energy between the spin
down electron and the spin wave, in relation to the energy of the
peak of the LCP spectrum obtained in the limit of small $d$. In
addition, we also showed that the recombination of the
skyrmion-hole state might be responsible for the low-energy tail
experimentally observed in the RCP spectrum.

Despite the fact that our model is not able to determine the
energy of the photoexcited ground states, we can analytically
calculate the spectral lines and also study a model for the
quantum Hall skyrmion.

From our analysis, we can conclude that the photoluminescence
experiment also corroborates the existence of a composite
excitation in the 2DEG at $\nu = 1$ as the quantum Hall skyrmion.

\begin{acknowledgments}
RLD and AOC kindly acknowledge Funda\c{c}\~ao de Amparo \`a Pesquisa de
S\~ao Paulo (FAPESP) for the financial support and AOC also
acknowledges the support from the Conselho Nacional de Desenvolvimento
Cient\'{\i}fico e Tecnol\'ogico (CNPq).
\end{acknowledgments}

\appendix

\section{\label{funcaog} The $G_{m,m'}(x)$ function}

The function $G_{m,m'}(x)$ is related to the matrix element of the
operator $\exp(-i\mathbf{q}\cdot\mathbf{r})$ in the lowest Landau
level. It is defined as \cite{doretto}
\begin{widetext}
\begin{eqnarray}
\label{gdef2}
\nonumber
G_{m,m'}(lq) &=&
\theta(m'-m)\sqrt{\frac{m!}{m'!}}(\frac{-ilq^*}{\sqrt{2}})^{m'-m}
        L^{m'-m}_{m}(\frac{|lq|^2}{2})
%\\ \nonumber && \\ \nonumber
%    &+&
        +\theta(m-m')\sqrt{\frac{m'!}{m!}}(\frac{-ilq}{\sqrt{2}})^{m-m'}
        L^{m-m'}_{m'}(\frac{|lq|^2}{2}),
\end{eqnarray}
where $L^{m-m'}_{m'}(x)$ is the generalized Laguerre polynomial.

%Among the several properties of this function (a complete list is
%presented in Ref. \cite{doretto}), we mention the one related to the sum
%of the product of two functions,
%
%\begin{eqnarray}
%\nonumber
%\sum_{l}G_{m,l}(lq)G_{l,m'}(lk) &=& \sum_{l}
%\langle m|\exp(-il/\sqrt{2}qb^{\dagger})\exp(-il/\sqrt{2}q^*b)|l\rangle
%\\ \nonumber &&\\ \nonumber
%&*&
%   \langle l|exp(-il/\sqrt{2}kb^{\dagger})\exp(-il/\sqrt{2}k^*b)|m'\rangle
%\\ \nonumber \\ \label{propg3}
%&=& \exp\left(\frac{-l^2q^*k}{2}\right)G_{m,m'}(lq + lk).
%\end{eqnarray}

\section{\label{cc} bosonic form of the product of fermionic operators}

Using the bosonization method for the two-dimensional electron gas
at $\nu=1$ introduced in Ref. \onlinecite{doretto}, it is possible to show
that the product of fermionic
operators can be written in a bosonic language as
\begin{eqnarray}
\label{cc1}
c^{\dagger}_{m\;\downarrow}c_{m'\;\uparrow} &\equiv&
   \frac{1}{\sqrt{N_{\phi}}}\sum_{\mathbf{q}}e^{-|lq|^2/4}
   G_{m',m}(-lq)b^{\dagger}_{\mathbf{q}},
\\ \nonumber && \\ \label{cc2}
c^{\dagger}_{m\;\uparrow}c_{m'\;\downarrow} &=&
\frac{1}{\sqrt{N_{\phi}}}\sum_{\mathbf{k}}e^{-|lk|^2/4}G_{m',m}(lk)b_{\mathbf{k}}
-\sum_{\mathbf{p,q,k}}\frac{e^{-|lk|^2/4}}{N^{3/2}_{\phi}}
\cos\left(\frac{\mathbf{(k+p)}\wedge\mathbf{(p-q)}}{2}\right)
G_{m',m}(lk)b^{\dagger}_{\mathbf{k+p+q}}b_{\mathbf{p}}b_{\mathbf{q}},\;\;\;\;
\\ \nonumber && \\ \label{cc3}
c^{\dagger}_{m\;\uparrow}c_{m'\;\uparrow} &=& \delta_{m,m'}
  - \frac{1}{N_{\phi}}\sum_{\mathbf{k,q}}e^{-|l(k-q)|^2/4}  
e^{-i\mathbf{k}\wedge\mathbf{q}}G_{m',m}(lq-lk)b^{\dagger}_{\mathbf{k}}b_{\mathbf{q}},
\\ \nonumber && \\ \label{cc4}
c^{\dagger}_{m\;\downarrow}c_{m'\;\downarrow} &=&
  \frac{1}{N_{\phi}}\sum_{\mathbf{k,q}}e^{-|l(k-q)|^2/4}  
e^{i\mathbf{k}\wedge\mathbf{q}}G_{m',m}(lq-lk)b^{\dagger}_{\mathbf{k}}b_{\mathbf{q}},
\end{eqnarray}
where 
$\mathbf{k}\wedge\mathbf{q}=l^2\hat{z}\cdot(\mathbf{k}\times\mathbf{q})$.
The above expressions are quite similar to the ones derived in
Ref. \onlinecite{doretto} for the spin density operators.

\section{\label{apendice:rcp} The $\mathcal{F}_{RCP}(t)$  function for
         the skyrmion-hole initial state}

As $[b_{\mathbf{q}},c_{m\;\downarrow}]=0$,
the function $\mathcal{F}_{RCP}(t)$, defined in Eq. (\ref{taxarcp}), can be written as
\begin{eqnarray}
\nonumber
\mathcal{F}_{RCP}(t)&=& \langle FM|\prod_{i=1}^n b_\mathbf{q_i}      
\left(\delta_{m,0}-c^{\dagger}_{0\;\downarrow}c_{m\;\downarrow}\right)
        e^{-i\mathcal{H}_ft}
\left(\delta_{m,0}-c^{\dagger}_{m\;\downarrow}c_{0\;\downarrow}\right)
        \prod_{i=1}^n b^{\dagger}_\mathbf{q_i} |FM\rangle \nonumber \\
&& \nonumber \\ \label{frcp1}
        &=& \sum_{m=0}^4 \mathcal{F}^{(m)}_{RCP}(t).
\end{eqnarray}
Let us analyze each of the four terms separately.

The first term is simply given by
\begin{eqnarray}
\nonumber
\mathcal{F}^{(1)}_{RCP}(t) &=& \delta_{m,0}\langle FM|
        \prod_{i=1}^n b_\mathbf{q_i}
        e^{-i\mathcal{H}_ft}
        \prod_{i=1}^n b^{\dagger}_\mathbf{q_i}|FM\rangle. 
\end{eqnarray}
Writing $\mathcal{H}_f$ as in Eq. (\ref{ho}) and considering the long
wavelength limit of the boson dispersion relation $w_\mathbf{q}$
[Eq. (\ref{expansaorpa})], with the aid of the expression (\ref{auxiliar}), the first
term reduces to
\begin{eqnarray}
\label{frcp1}
\mathcal{F}^{(1)}_{RCP}(t) &=& \delta_{m,0}\exp\left(-it(\epsilon + E_{FM})\right).
\end{eqnarray}
For the second one, we have
\begin{eqnarray}
\nonumber
\mathcal{F}^{(2)}_{RCP}(t) &=& -\delta_{m,0}\langle FM|
        \prod_{i=1}^n b_\mathbf{q_i}
        e^{-i\mathcal{H}_ft}
        c^{\dagger}_{0\;\downarrow}c_{0\;\downarrow}
        \prod_{i=1}^n b^{\dagger}_\mathbf{q_i}|FM\rangle 
\\ \nonumber && \\ \nonumber
        &=& -\frac{1}{N_\phi}\delta_{m,0}\sum_{\mathbf{k,k'}}e^{-|l(k-k')|^2/4}  
            e^{i\mathbf{k}\wedge\mathbf{k'}}
            \langle FM|\prod_{i=1}^n b_\mathbf{q_i}
            e^{-i\mathcal{H}_ft}
            b^{\dagger}_\mathbf{k}b_\mathbf{k'}
            \prod_{i=1}^n b^{\dagger}_\mathbf{q_i}|FM\rangle.
\end{eqnarray}
Here, in the second step, we use the fact that the bosonic form of the
operator $c^{\dagger}_{0\;\downarrow}c_{0\;\downarrow}$ is a
linear combination of the product
$b^{\dagger}_{\mathbf{k}}b_{\mathbf{k'}}$ 
[see Eq. (\ref{cc4})]. Reordering the bosonic operators, we end up
with 
\begin{eqnarray}
\mathcal{F}^{(2)}_{RCP}(t) &=& 
      -\frac{n}{N_\phi}\delta_{m,0}\exp\left(-it(\epsilon + E_{FM})\right),
\label{frcp2}
\end{eqnarray}
where $n$ is the number of spin-waves in the state (\ref{skyrmion}).
The third term is similar to the second one and therefore the former
is equal to (\ref{frcp2}).

Finally, the last term is given by
\begin{eqnarray}
\nonumber
\mathcal{F}^{(4)}_{RCP}(t) &=& \langle FM|
        \prod_{i=1}^n b_\mathbf{q_i}c^{\dagger}_{0\;\downarrow}c_{m\;\downarrow}
        e^{-i\mathcal{H}_ft}
        c^{\dagger}_{m\;\downarrow}c_{0\;\downarrow}
        \prod_{i=1}^n b^{\dagger}_\mathbf{q_i}|FM\rangle
\end{eqnarray}
Again, substituting Eqs. (\ref{cc1}) and (\ref{cc2}) in the above
expression, choosing $m=0$ and reordering the bosonic operators, we
have
\begin{eqnarray}
\nonumber
\mathcal{F}^{(4)}_{RCP}(t) &=&
  \frac{1}{N^2_\phi}\sum_{j=1}^n\sum_\mathbf{p}e^{-|l(\mathbf{p - q_j})|^2/2}  
  \exp\left(-it(\epsilon + E_{FM} - w_\mathbf{q_j} + w_\mathbf{p})\right)
\\ \nonumber && \\
  && + \frac{1}{N^2_\phi}\sum_{j=1}^n\sum_{l\not= j}\left[ e^{-|l(\mathbf{q_j - q_l})|^2/2}  
  \exp\left(-it(\epsilon + E_{FM} - w_\mathbf{q_j} + w_\mathbf{q_l})\right)
     + \exp\left(-it(\epsilon + E_{FM})\right)\right].
\end{eqnarray}
Substituting $\mathbf{p}\rightarrow\mathbf{p+q_j}$ in the first term
of the above expression, changing the sum over momenta into an integral and rescaling
the momentum $p \rightarrow lp$, we find that
\begin{eqnarray}
\nonumber
\mathcal{F}^{(4)}_{RCP}(t) &=&
   -\frac{2i}{N_\phi}\sum_{j=1}^n \exp\left(-it(\epsilon + E_{FM})\right)
   \frac{1}{t\epsilon_B - 2i}
   \exp\left(\frac{-i(lq_jt\epsilon_B)^2}{8i-4t\epsilon_B}\right)
\\ \nonumber && \\ \nonumber
  && + \frac{1}{N^2_\phi}\sum_{j=1}^n\sum_{l\not= j}\left[ e^{-|l(\mathbf{q_j - q_l})|^2/2}  
  \exp\left(-it(\epsilon + E_{FM} - w_\mathbf{q_j} + w_\mathbf{q_l})\right)\right]
\\ \nonumber && \\ \label{frcp4}
  && + \frac{1}{N^2_\phi}n(n-1)\exp\left(-it(\epsilon + E_{FM})\right).
\end{eqnarray}
As we assume that the bosons momenta are $|lq_j|\ll 1$, we can expand
the first term of Eq. (\ref{frcp4}), i.e.,
\begin{equation}
\exp(\frac{-i(lq_jt\epsilon_B)^2}{8i-4t\epsilon_B}) \approx 1 + \mathcal{O}(|lq_j|^2).
\end{equation}
Therefore, from the expressions (\ref{frcp1}), (\ref{frcp2}) and
(\ref{frcp4}), we find that the function $\mathcal{F}_{RCP}(t)$ is
given by the equation (\ref{frcp}).
\end{widetext}

\bibliographystyle{apsrev}

\begin{thebibliography}{99}
\bibitem{zyun} Z. F. Ezawa, {\it Quantum Hall effects - field theoretical 
                  approach and related
                  topics} (World Scientific, Singapore, 2000).

\bibitem{perspectives} {\it Perspectives in Quantum Hall Effects},
  edited by S. Das Sarma and A. Pinczuk (Wiley, New York, 1997). 

\bibitem{kallin} C. Kallin and B. I. Halperin, Phys. Rev. B {\bf 30},
                 5655 (1984).

\bibitem{sondhi} S. L. Sondhi, A. Karlhede, S. A. Kivelson and E. H. Rezayi,
                 Phys. Rev.  {\bf 47}, 16419 (1993).

\bibitem{plentz}  F. Plentz, D. Heiman, L. N. Pfeiffer and K. W. West,
                 Phys. Rev. B {\bf57}, 1370 (1998).

\bibitem{cooper} N. R. Cooper and D. B. Chklovskii, Phys. Rev. B 
                 {\bf 55}, 2436 (1997).

\bibitem{osborne}J. L. Osborne, A. J. Shields, M. Y. Simmons, N. R. Cooper,
                 D. A. Ritchie and M. Pepper, Phys. Rev. B
                 {\bf 58}, R4227 (1998).

\bibitem{schotte}K. D. Schotte and U. Schotte, Phys. Rev.
                 {\bf 182}, 479 (1969).

\bibitem{harry2} H. Westfahl Jr., A. O. Caldeira, D. Baeriswyl and
                 E. Miranda, Phys. Rev. Lett. {\bf 80}, 2953
                 (1998).

\bibitem{doretto} R. L. Doretto, A. O. Caldeira and S. M. Girvin,
                  Phys. Rev. B {\bf 71}, 045339 (2005).

\bibitem{grads} I.S. Gradshteyn and I.M. Ryzhik, \emph{Table of integrals, 
                  series and products}(Academic Press Inc., 1980), 4th.

\bibitem{mahan}  G. Mahan, \emph{Many particle physics}
                 (Plenum, New York, 1981).

\bibitem{portengen} T. Portengen, J. R. Chapman, V. N. Nicopoulus and N. F. 
                    Johnson, Int. J. Mod. Phys. B {\bf 12}, 1 (1998).

\bibitem{wojs} A. W\'ojs and J. J. Quinn, cond-mat/0308410; A. W\'ojs
  and J. J. Quinn, cond-mat/0308411.

\bibitem{palacios} J. J. Palacios and H. A. Fertig, Phys. Rev. Lett.
                   {\bf 79}, 471 (1997).

\bibitem{oaknin}  J. H. Oaknin, B. Paredes and C. Tejedor, Phys. Rev. B 
                  {\bf 58}, 13028 (1998).


\end{thebibliography}

\end{document}